\documentclass[fleqn,usenatbib]{mnras}

\usepackage{newtxtext,newtxmath}

\usepackage{graphics}

\usepackage[T1]{fontenc}

\DeclareRobustCommand{\VAN}[3]{#2}
\let\VANthebibliography\thebibliography
\def\thebibliography{\DeclareRobustCommand{\VAN}[3]{##3}\VANthebibliography}

\usepackage{graphicx}	
\usepackage{amsmath}	
\usepackage{parskip}
\usepackage{xcolor}
\usepackage{dcolumn}

\usepackage{epstopdf}
\usepackage{hyperref}

\graphicspath{{figures/}}

\newcommand{\CTWO}{${\mathrm C}_{2}$}
\newcommand{\CTHREE}{${\mathrm C}_{3}$}
\newcommand{\CSIXTY}{${\mathrm C}_{60}$\ }
\newcommand{\CTHCYH}{${\mathrm c-}{\mathrm C}_{3}{\mathrm H}$}
\newcommand{\CTHHTH}{${\mathrm C}_{3}{\mathrm H}_{3}$}
\newcommand{\XIICTHREE}{$^{12}{\mathrm C}_{3}$}
\newcommand{\XIIICTHREE}{$^{13}{\mathrm C}_{3}$}
\newcommand{\XIIXIII}{$^{12}{\mathrm C}_{3}/^{13}{\mathrm C}_{3}$}
\newcommand{\cmi}{${\rm cm}^{-1}$}
\newcommand{\OMC}{${\mathrm O} - {\mathrm C}$}
\newcommand{\CCC}{$^{12}{\mathrm C}_3$}

\newcommand{\CSC}{$^{12}{\mathrm C}^{13}{\mathrm C}^{12}{\mathrm C}$}
\newcommand{\CCS}{$^{12}{\mathrm C}^{12}{\mathrm C}^{13}{\mathrm C}$}
\newcommand{\SCS}{$^{13}{\mathrm C}^{12}{\mathrm C}^{13}{\mathrm C}$}
\newcommand{\SSC}{$^{13}{\mathrm C}^{13}{\mathrm C}^{12}{\mathrm C}$}
\newcommand{\SSS}{$^{13}{\mathrm C}^{13}{\mathrm C}^{13}{\mathrm C}$}
\newcommand{\OOOO}{$(0,0^{0},0)$}
\newcommand{\OIIO}{$(0,1^{1},0)$}
\newcommand{\OOOI}{$(0,0^{0},1)$}
\newcommand{\OIII}{$(0,1^{1},1)$}
\newcommand{\IOOI}{$(1,0^{0},1)$}
\newcommand{\IIII}{$(1,1^{1},1)$}

\newcommand{\ai}{\textit{ab initio}}

\newcommand{\cm}{cm$^{-1}$}

\usepackage{color}

\newcommand{\name}{``AtLast''}

\newcolumntype{d}[1]{D{.}{.}{#1}}
\title[ExoMol Line Lists -- LXII. \CTHREE]{ExoMol Line Lists -- LXII: Ro-Vibrational Energy Levels and \\ Line-Strengths for the Propadienediylidene (\CTHREE) in its Ground Electronic State}

\author[A. E. Lynas-Gray et al.]{
A. E. Lynas-Gray$^{1,2,3}$,
O. L. Polyansky$^{1,4}$,
J. Tennyson$^{1}$\thanks{Corresponding Author E-mail: j.tennyson@ucl.ac.uk},
S. N. Yurchenko$^{1}$ \newauthor
and
N. F. Zobov$^{4}$
\\ \\
$^{1}$Department of Physics and Astronomy, University College London, Gower Street, London WC1E 6BT, UK\\
$^{2}$Department of Physics, University of Oxford, Keble Road, Oxford OX1 3RH, UK\\
$^{3}$Department of Physics and Astronomy, University of the Western Cape, Bellville 7535, South Africa\\
$^{4}$Institute of Applied Physics, Russian Academy of Science, Ulyanov Street 46, Nizhnii Novgorod 603950, Russia
}

\date{Accepted 2024 October 22. Received 2024 October 21; in original form 2024 August 6}

\pubyear{2024}

\begin{document}
\label{firstpage}
\pagerange{\pageref{firstpage}--\pageref{lastpage}}
\maketitle

\begin{abstract}
Improved opacities are needed for modelling the atmospheres and evolution of cool carbon-rich stars and extra-solar planets;
in particular, contributions made by the astrophysically important
propadienediylidene (\CTHREE) molecule need, at a minimum, to be
determined using a line list which includes all significant transitions
in the energy range of interest.  We report variational calculations
giving ro-vibrational energy levels and corresponding line-strengths
for \CCC, \CSC\ and \CCS. In the \CCC\ case we obtain 2~166~503 ro-vibrational
state energies $\leqslant 20000$~\cmi\ for the electronic
$\tilde{X}{\,}^{1}{\Sigma_{\rm g}}^{+}$ ground-state.
Comparison with experiment indicates a maximum error of
$\pm 0.03$~\cmi\ in calculated positions of lines involving an upper
state energy $\lessapprox 4000$~\cmi.  For lines with upper state
energies $\gtrapprox 4000$~\cmi\ to have comparable line-position
accuracies, conical intersections would need to be accounted for in an
adopted potential energy surface. Line lists and associated opacities are provided in
the ExoMol Database (\url{http://www.exomol.com}). 
  \\
\end{abstract}
%


\setlength{\parskip}{0.3cm}

\section{Introduction}

Interest in propadienediylidene (tricarbon, hereinafter \CTHREE)
follows the demonstration by \citet{1951ApJ...114..466D} that the
molecular band near 4050~\AA\ in cometary spectra may be
attributable to it.  First observed by \citet{1881RSPS...33....1H},
the $~$4050~\AA\ band is also seen in carbon star spectra
\citep{1948ApJ...108..453M} and work by
\citeauthor{1951ApJ...114..466D}
allowed \citet{1953MNRAS.113..571S} to identify \CTHREE\ as the carrier.
A detection of \CTHREE\ is to be expected as
\citeauthor{1954JChPh..22..126H}'s \citeyearpar{1954JChPh..22..126H}
determination of the dissociation energy (the energy required to
break a C--C bond) of 160~kcal/mol (6.938~eV) led
\citet{1964PJAB...40...99T} to demonstrate that both
\CTWO\ and \CTHREE\ achieve maximum number densities in conditions
prevailing in some carbon star atmospheres at temperatures
$\lessapprox 2800$~K.

\citet{1965ApJ...142...45G} reproduced the ${\rm C}_3$ 4050~\AA\ band in
fluorescence and absorption using flash photolysis of diazomethane
mixed with an excess (100:1) of nitrogen.  The electronic transition
$\tilde{A}{\,}^{1}{\Pi}_{\rm u} - \tilde{X}{\,}^{1}{\Sigma}_{\rm g}^{+}$
was found to produce the
${\rm C}_3$ 4050~\AA\ band.  Following their rotational analysis,
\citeauthor{1965ApJ...142...45G} established through a vibrational
analysis that there are two low lying $\tilde{X}{\,}^{1}{\Sigma_{\rm g}}^{+}$ levels
at 132~${\rm cm}^{-1}$ and 286~${\rm cm}^{-1}$ which have to be
respectively assigned as $2{\nu}_2$ and $4{\nu}_2$, leading to the
discovery of the surprisingly low-order bending mode
${\nu}_2 \approx 64\,{\rm cm}^{-1}$.

\citet{2001ApJ...551L.181G} exploited the \CTHREE\ low-order bending mode
to establish its presence in the interstellar medium along the
line-of-sight towards Sgr B2(M).  Reactions involving \CTHREE\ in
the interstellar medium are understood to determine carbon chain
abundances in dark clouds \citep{2014MNRAS.437..930L}.  Because \CTWO\
and \CTHREE\ are obvious building blocks for larger carbon molecules
such as \CSIXTY, \citet{2024A+A...681A...6F} located \CTHREE\ in a
further twenty-seven lines-of-sight for follow-up observations.

\citet{2013A+A...552A.132H} revised \CTHREE-hydrocarbon models of
Titan's atmosphere, finding \CTHREE, \CTHCYH\ and \CTHHTH\ may be
detectable. Results from high dispersion optical spectroscopy
\citep{2024P+SS..24005836R} are consistent with a \CTHREE\ column
density of $10^{13}\,{\rm cm}^2$ in the upper atmosphere of Titan.
An updated line list enabled \citet{2019ApJ...881L..33L} to detect
propadiene
(${\mathrm C}{\mathrm H}_2{\mathrm C}{\mathrm C}{\mathrm H}_2$)
in the atmosphere of Titan using infrared high dispersion spectra;
their spectra could also be used to search for \CTHREE\ were an
adequate line list available.

\cite{1996PASP..108..225C} reviewed R~Corona~Borealis (RCrB) stars whose atmospheric
compositions are 99\% helium and 1\% carbon (by numbers); their light
curves exhibit sharp and irregular decreases in brightness of roughly
6-8 magnitudes, the original magnitude being recovered only after a
slow increase in brightness which follows.  Soot formation and
dissipation in RCrB atmospheres is understood to be the cause of the
characteristic RCrB light-curves and it was therefore of some interest
to note that laboratory studies of \CTHREE\ show it to be the dominant
vapour species over graphite above $\approx 2000$~K and identify it as a
potential precursor for soot formation in flames
\citep{2010MolPh.108.1013G}.  \cite{1990MNRAS.244...29K} identify the
4050~\AA\ \CTHREE\ band in a RCrB spectrum obtained at maximum light
and find it to have disappeared as minimum light is approached,
suggesting that \CTHREE\ is now absent and soot formation has taken
place. 


Experiments by \citet{1964JChPh..40.1299W} and
\citet{1964JChPh..40.1305W} respectively located the \CTHREE\ ground state
asymmetric and symmetric stretches at 2040~${\rm cm}^{-1}$ and
1230~${\rm cm}^{-1}$.  \citet{1973IAUS...52..517G} then argued that
\CTHREE\ should be included as a contributor to a broad absorption feature seen
in carbon star spectra at ${\sim}5\,\mu{\rm m}$ and that it contributes significant
opacity at this frequency.
\citet{1975ApJ...202..839T} studied the vibrational
spectrum of \CTHREE\ in the ${\sim}5\,\mu{\rm m}$ region, using a carbon
tube furnace at 3100~K; they found a broad unresolved feature at
${\sim}2000\,{\rm cm}^{-1}$ which is ${\sim}370\,{\rm cm}^{-1}$
wide and unambiguously identified with the asymmetric stretch
ro-vibrational band of \CTHREE.

\citet{2004A+A...422..289G} compared phase-dependent observed and
synthetic carbon star spectra, the calculations being based on
dynamical model stellar atmospheres by \citet{2003A+A...399..589H}.
Of specific interest was the comparison with the \CTHREE\
${\sim}5.1\,\mu{\rm m}$ feature which \citeauthor{2004A+A...422..289G}
had difficulty in modelling.  Among possible explanations in need of
investigation is \citeauthor{2003A+A...399..589H}'s use of opacities
from the \citet{1997IAUS..178..441J} database which in the \CTHREE\ case
are based on the line list by \citet{1989ApJ...343..554J}.
\citet{2018Atoms...6...26T} present an ExoMol
\citep{2012MNRAS.425...21T} atlas of molecular opacities and
note that by modern standards
the \citet{1989ApJ...343..554J} \CTHREE\ line list cannot be
considered reliable, given improved methods and routine tuning of
calculated line-frequencies to experimental data.

\citet{2015JChPh.143g4302R} published a global \ai\ potential energy surface
(PES) obtained using the MRCI/AVTZ//FVCAS/AVTZ level of theory,  which correctly describes the overall topology of the \CTHREE\
electronic ground state; this includes the all important conical
intersections.  Quite good agreement is achieved for pure
bending modes but the stretching modes are not calculated with the
accuracy required for a high quality line list.
However, \citet{2016JChPh.144d4307S} calculated an accurate local
(near equilibrium) \ai\ PES using the fc-CCSD(T*)-F12b/aug-cc-pV5Z level of theory which reproduces all
then available rotational-vibrational term energies to better than 1~\cmi; they achieve this by taking special care with potential energy
convergence relative to high-order correlation effects,
core-variance correlation, basis set size and scalar relativity.
Moreover, the rotational constants exhibit relative errors of not more
than 0.01\%. Regardless of a decent \ai\ quality, this is still not sufficiently accurate for line list applications. We have therefore constructed a new PES for C$_3$ by refining an \ai\ PES to experimentally derived ro-vibrational energies by \citet{jt915} and chosen the \citeauthor{2016JChPh.144d4307S}
PES as a starting point for these refinements. 

\citet{2016JChPh.144d4307S} reported an \ai\ electric dipole moment
surface (EDMF) of \CTHREE\ using the fc-CCSD(T*)-F12b/AV5Z level of theory which we adopt in our line list intensity calculations. 
Note that this allows us to provide absolute transition intensities for all transitions that we consider,
something missing from all experimental studies reported up until now.

\citet{jt939} summarised progress made with the ExoMol
project up to April 2024, and the
provision of a database from which line lists and associated data
may be retrieved.  Examples of available line lists include alkali metal hydroxides \citep[KOH and NaOH,][]{2021MNRAS.502.1128O},
silicon monoxide \citep{2022MNRAS.510..903Y}, calcium and magnesium
monohydride \citep[CaH and MgH,][]{2022MNRAS.511.5448O},
silicon mononitrate \citep[SiN,][]{2022MNRAS.516.1158S},
calcium monohydroxide \citep[CaOH,][]{2022MNRAS.516.3995O},
${\rm H_3}^+$ isotopologues \citep[${\rm H_3}^+$, ${\rm H_2D}^+$,
${\rm HD_2}^+$ and ${\rm D_3}^+$,][]{2023MNRAS.519.6333B},
thioformaldehyde \citep{2023MNRAS.520.1997M},
aluminium monochloride \citep[AlCl,][]{2023MNRAS.520.5183Y},
lithium hydroxide \citep[LiOH,][]{2024MNRAS.527..731O},
yttrium oxide \citep[YO,][]{2024MNRAS.527.4899Y},
sulphur monoxide \citep[SO,][]{2024MNRAS.527.6675B},
aluminium hydride and aluminium deuteride \citep[AlH and AlD,][]
{2024MNRAS.527.9736Y} and the
methylidene cation \citep[$\rm{CH}^+$,][]{2024MNRAS.52710726P}.

Reported in the present paper are calculations which produce ExoMol
database line lists for \CCC, \CSC\ and \CCS; the choice of isotopologues having been
influenced by the detection  by \citet{2020A+A...633A.120G} of \CSC\ and
\CCS\ in the direction of Sgr~B2(M). \citet{2016JChPh.145w4302B}
mention an earlier failure to find interstellar \CCS\ using the
Herschel space telescope, assuming a $\nu_2$ lowest bending mode of
60.747~\cmi for \CCS\ as \citet{2013JPCA..117.3332K} reported.
Laboratory experiments by \citet{2016JChPh.145w4302B} identified the
$\nu_2 = 60.747$~\cmi absorption as belong to \SSC; as this was not observed
we concluded that there was no immediate need for isotopolgue
line lists where more than one $^{12}{\mathrm C}$ has been substituted by $^{13}{\mathrm C}$.

The summary of early work on \CTHREE\ presented above is necessarily
brief and focussed on astronomical spectroscopy. For more details,
including laboratory studies, see \citet{2023JMoSp.39111734M} and
\citet{jt915}.

\begin{table}
\begin{center}
\caption{Comparison Between Calculated (TROVE) and Observed (MARVEL) State Energies
Used in Potential Energy Surface Fitting}
\label{tabstates}
\begin{tabular}{ccccccd{5.4}d{5.4}d{7.4}}
\hline
       &        &        &        &    &     &   \multicolumn{3}{c}{ }          \\
 $v_1$ &  $v_2$ & $\ell$ &  $v_3$ &  J &  p  &   \multicolumn{3}{c}{State Energy$\,$(\cmi)} \\
       &        &        &        &    &     &   \multicolumn{1}{c}{TROVE} & \multicolumn{1}{c}{MARVEL} & \multicolumn{1}{c}{$\:$\OMC}\\
       &        &        &        &    &     &   \multicolumn{3}{c}{ }          \\
\hline
       &        &        &        &    &     &   \multicolumn{3}{c}{ }          \\
   0   &   0    &    0   &    0   &  0 &  e  &    0.000000 &   0.0           &  0.00000 \\
   0   &   1    &    1   &    0   &  1 &  e  &   63.853305 &   63.8533045(4) & -0.00000 \\
   0   &   2    &    0   &    0   &  0 &  e  &  132.795482 &  132.795(6)     & -0.00048 \\
   0   &   2    &    2   &    0   &  2 &  e  &  133.938774 &  133.939(6)     & +0.00023 \\
   0   &   3    &    1   &    0   &  1 &  e  &  207.872923 &  207.873(6)     & +0.00008 \\
   0   &   4    &    0   &    0   &  0 &  e  &  286.557525 &  286.558(6)     & +0.00047 \\
   0   &   4    &    2   &    0   &  2 &  e  &  288.155899 &  288.156(6)     & +0.00010 \\
   0   &   4    &    4   &    0   &  4 &  e  &  291.042029 &  291.042(16)    & -0.00003 \\
   0   &   5    &    3   &    0   &  3 &  e  &  373.462909 &  373.463(5)     & +0.00009 \\
   0   &   5    &    5   &    0   &  5 &  e  &  377.523167 &  377.523(5)     & -0.00017 \\
   1   &   0    &    0   &    0   &  0 &  e  & 1224.524735 & 1224.52(1)      & -0.00473 \\
   0   &   1    &    1   &    1   &  1 &  f  & 2078.957902 & 2078.957(3)     & -0.00090 \\
   0   &   2    &    2   &    1   &  2 &  f  & 2128.302969 & 2128.302(8)     & -0.00097 \\
   0   &   4    &    4   &    1   &  4 &  f  & 2251.089429 & 2251.089(15)    & -0.00043 \\
   0   &   5    &    5   &    1   &  5 &  f  & 2322.740567 & 2322.741(7)     & +0.00043 \\
   0   &   5    &    3   &    1   &  3 &  f  & 2329.286309 & 2329.286(9)     & -0.00031 \\
   2   &   0    &    0   &    0   &  0 &  e  & 2436.126153 & 2436.1(6)       & -0.02615 \\
   1   &   1    &    1   &    1   &  1 &  f  & 3330.949595 & 3330.9496(5)    & +0.00000 \\
   1   &   0    &    0   &    2   &  0 &  e  & 5268.399364 & 5268.399(2)     & -0.00036 \\
   2   &   0    &    0   &    2   &  0 &  e  & 6460.662620 & 6460.663(2)     & +0.00038 \\
   3   &   0    &    0   &    2   &  0 &  e  & 7635.526541 & 7635.526(18)    & -0.00054 \\
   4   &   0    &    0   &    2   &  0 &  e  & 8799.534353 & 8799.5(6)       & -0.03435 \\
\hline
\end{tabular}
\end{center}
\end{table}

\begin{table*}
\begin{center}
\caption{TROVE Input Parameters and Numbers of Transitions Calculated}
\label{trveio}
\begin{tabular}{ccccccr}
\hline
             &                 &         &        &          &           &              \\
Isotopologue & Number of       &  Energy & $J$      &  $\ell$  & Number of & Number of    \\
             & Basis Functions & Maximum &Range   &  Range   &  States   & Transitions  \\
             &                 &  [\cmi] &        &          &           &              \\
             &                 &         &        &          &           &              \\
\hline
             &                 &         &        &          &           &              \\
\CCC         &       56        &  20000  &$0-155$ & $0-12$   & 2166503   &  5481690507  \\
\CSC         &       56        &  20000  &$0-155$ & $0-12$   & 2282841   &  6071530477  \\
\CCS         &       45        &  17000  &$0-150$ & $0-10$   & 2442205   & 14503868150  \\
             &                 &         &        &          &           &              \\
\hline
\end{tabular}
\end{center}
\end{table*}

\section{Ro-vibrational calculations}

\citet{2007JMoSp.245..126Y} published  the variational methodology and associated program (TROVE)   for general 
calculations of ro-vibrational energies for an arbitrary small or medium-sized polyatomic. For triatomics, TROVE  can use an exact kinetic energy operator (KEO) as developed by \citet{20YuMexx} for quasi-linear molecules, adopted here for our \CTHREE\ line list calculations using the PES and EDMF discussed below. We follow closely the calculation procedure described in detail by \citet{20YuMeFr}, where TROVE was used to compute a ro-vibrational line list for CO$_2$. The KEO is built using the bisector frame and valence coordinates $r_{1}$, $r_2$ (bond lengths) and $\rho = \pi -\alpha$, where $\alpha$ is a bond angle.

TROVE uses numerically constructed 1D primitive basis functions optimised for a given PES. The stretching basis functions $\phi_{v_1}(r_1)$ and $\phi_{v_2}(r_2)$ are generated using a Numerov-Cooley procedure~\citep{24Numerov.method,61Cooley.method}, while the bending basis functions $\phi_{v_3}(\rho)$ are obtained by solving the corresponding Schr\"{o}dinger equation using associated Laguerre polynomials, see \citet{20YuMexx}. For \CTHREE, the basis set was limited by the following conditions
\begin{equation}
 v_1+v_2+v_3 \le 56, 
\end{equation}
and $v_1\le 30$, $v_2\le 30$ and $v_3\le 56$. 
Extra bending functions are needed to allow for the low frequency of the bending mode and the large range of angles (approaching 90$^\circ$) sampled
by the excited bending states we consider here.
The primitive basis functions are then further improved through a two-step contraction procedure. The final ro-vibrational basis functions are formed as symmetrically adapted products (see \citet{17YuYaOv}  for details)
\begin{equation}
    \Phi_{n,K}^{J,\Gamma} = \Phi_{n}^{(J=0),\Gamma_{\rm vib}} |J,K,\tau\rangle,  
\label{e:Phi}\end{equation}
where $\Phi_{n}^{(J=0),\Gamma_{\rm vib}}$ is an eigenfunction of the vibrational ($J=0$) Schr\"{o}dinger equation and $|J,K,\tau\rangle$ is a symmetry adapted (Wang) rigid rotor wavefunction with $\tau$ as its parity. 
 Here $n$ corresponds to the maximal value of the rotational quantum number $k$ (projection of the rotational angular momentum on the molecular axis $z$) in the ro-vibrational basis $n=10$.
In Eq.~(\ref{e:Phi}), $\Gamma$ and $\Gamma_{\rm vib}$ are the total and vibrational basis symmetries, respectively. For the two symmetric isotopologues \CCC\ and \CSC, the   C$_{2v}$(M) molecular symmetry group \citep{98BuJe.method} is used to classify the ro-vibrational eigen-solutions  spanning four irreducible representations $A_1$, $A_2$, $B_1$ and $B_2$. 
For the asymmetric isotopologue \CCS, the C$_{s}$(M) molecular symmetry group is used for the final ro-vibrational states. 
Internally, we employed  the so-called artificial extended molecular symmetry group C$_{nv}$(AEM) \citep{21MeYuJe,jt951}  to classify the rotational and bending basis functions, which helped reduce the memory requirement.

\section{Potential Energy and Dipole Moment Surfaces}

\citet{jt915} extracted observed transition wavenumbers
within and between the $\tilde{X}{\,}^{1}{\Sigma_{\rm g}}^{+}$
and $\tilde{A}{\,}^{1}{\Pi_{\rm u}}$ \CTHREE\ states from twenty-one
publications, subjecting these to a Measured Active
Rotational-Vibrational Energy Levels
\citep[MARVEL,][]{2007JMoSp.245..115F,2024JQSRT.31608902T} analysis which yields 1887
empirical energy levels.  We refined the
\citeauthor{2016JChPh.144d4307S} PES using those
$\tilde{X}{\,}^{1}{\Sigma_{\rm g}}^{+}$ corrected energy
levels \citeauthor{jt915} lists in his table~3,
with the exception of those at 4081.9, 4199.9, 4333.9 and 4486.7~\cmi\ for
which the energy uncertainties were comparatively high at
0.5~\cmi.

PES refinement followed the procedure described in detail by \citet{23Yurchenko} and used for many ExoMol line list calculations, including the very recent works on H$_2$CS \citep{2023MNRAS.520.1997M}, CH$_4$ \citep{24YuOwKe}, OCS \citep{jt943} and N$_2$O \citep{jt951}.  
The equilibrium bond  length and angle adopted were those \citeauthor{2016JChPh.144d4307S} use.  Following \citet{23Yurchenko}, our refined PES was also constrained to the \citeauthor{2016JChPh.144d4307S} PES at geometries not represented by MARVEL corrected state energies used in the PES refinement.
For the refinements we selected 276 ro-vibrational MARVEL term values for $J=0-10$, 20, 30, 40, 50, 60 with low uncertainties ($<0.1$~\cmi). These were reproduced with a root-mean-squares (rms) of 0.026~\cmi.

The quality of the fit is illustrated in Table~\ref{tabstates}, where we show TROVE predictions for 21 states compared to MARVEL energy levels for the $\tilde{X}{\,}^{1}{\Sigma_{\rm g}}^{+}$ \CTHREE\ state \citep[his table 3]{jt915}. \citeauthor{jt915}'s assignments are also replicated in Table~\ref{tabstates}: $v_1$ - symmetric stretch, $v_2$ - bend, $\ell$ - projection of $J$ on to the bond angle bisector, $v_3$ - asymmetric stretch, $J$ - rotational quantum number and $p$ - rotationless parity ($e$ even or $f$ odd). MARVEL energy uncertainties were placed in parentheses and give \citeauthor{jt915}'s estimate of the error in the least significant digits provided.  TROVE and MARVEL state energies were in agreement, even at $\sim $8800 \cmi\, to within experimental error in the latter; this was unexpected given the presence of conical intersections to be discussed below. Table~\ref{tabstates} residuals have a mean of $-(3.2 \pm 9.1) \times 10^{-3}$ \cmi.

\citet{2016JChPh.144d4307S} construct their electric dipole moment surface (EDMF) from 372 symmetry-unique nuclear configurations.  Local transformation to the Eckart coordinate system yields parallel and perpendicular components. Least squares fits gave
\citeauthor{2016JChPh.144d4307S} analytical expressions for the two components which we adopted for our line list calculations.

To summarise, our spectroscopic model consists of a refined PES and an \ai\ DMS of C$_3$, in its ground electronic state. We do not take into account any conical intersections or other couplings with other electronic states, see, e.g. \citet{2015JChPh.143g4302R}, nor any pre-dissociative effects -- C$_3$ has a high dissociation energy, $\sim 6.3$~eV \citep{2015JChPh.143g4302R}.


\section{Line list Calculations}


Table~\ref{trveio} lists TROVE input parameters selected,
as well as numbers of states and transitions calculated in each
case.  
Adopted carbon atom masses were 11.996709 and 13.00006335 Da; these are nuclear masses, with
core valence electron masses added.
Transitions included in our line lists were limited to those for which the upper state energy is less than $20000\,$~\cmi\ and the lower state energy is less than $10000\,$~\cmi. A lower state energy maximum of 10000~\cmi\  was selected in order to provide the sufficient temperature coverage for \CCS\  absorption spectra simulations, while the  upper energy thresholds is to maintain the line list completeness for the wavenumber range of 0--10000~\cmi. Rotation quantum number $J$ upper limits were  matched to the lower state energy threshold, above which no further population of ro-vibrational states were considered.

State energies in \CSC\ are slightly lower than corresponding state energies 
in \CCC, resulting in more being selected below a given maximum state energy, 
leading to more transitions being calculated in the \CSC\ case.   
A much larger number of transitions calculated for the \CCS\ isotopologue is a consequence of 
more parity-allowed transitions in the case of an asymmetric  molecule. Indeed, only even-parity 
states ($A_1$ and $A_2$)   exist for \CCC\ and \CSC\ due to zero nuclear spin of $^{12}$C 
and the Pauli principle, while the asymmetric species \CCS\ has no such restrictions 
and all states are present.

The line lists are provided in the ExoMol format \citep{jt939}, which consists of
a two file-type set, a states file and transitions files, augmented by a partition function.
The states file for \CTHREE\ contains the ro-vibrational energy term values, 
quantum numbers, life times, uncertainties and state degeneracies. Each state is 
labeled with a state ID, a counting number. The structure of the \CTHREE\ states file 
is illustrated in Table~\ref{t:states} where we show an extract from the \name\ line 
list for \CCC. We note that while the quantum number labels given for levels labelled
'Ma' for MARVELised can be regarded as secure; the quantum labels for other levels
are automatically generated by TROVE and represent only best estimates.

\begin{figure}
\centering
\includegraphics[width=0.95\columnwidth]{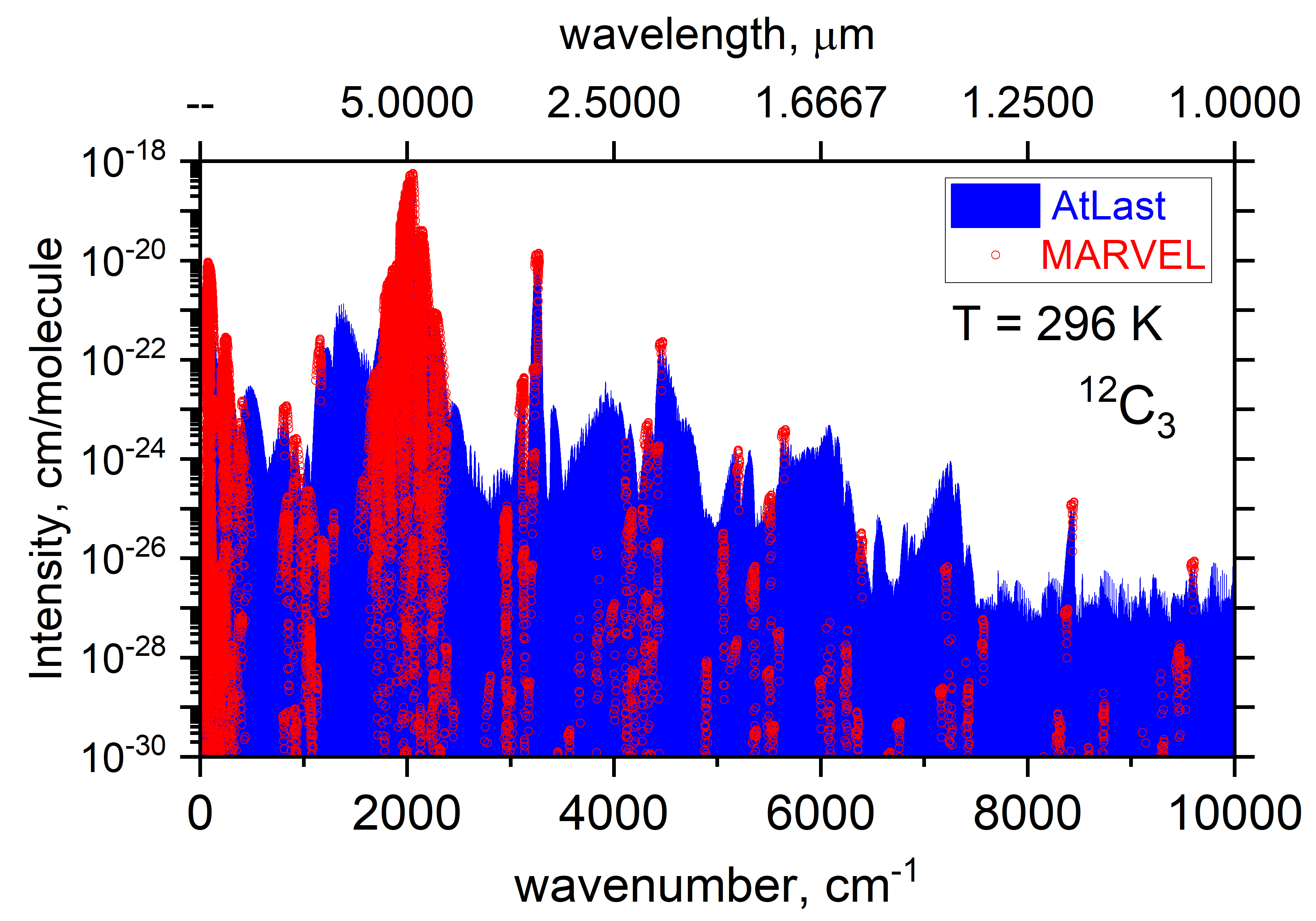}
\caption{A room temperature ($T=296$~K) ``stick spectrum'' \citep{2018A+A...614A.131Y} of \CCC\ computed using the \name\ line list, with the height of the sticks representing the transition line intensity (absorption coefficient) and their position representing the corresponding transition wavenumber. The MARVELised transitions  are indicated using red circles.}
\label{figMARVEL}
\end{figure}

Transitions files contain  Einstein~A-coefficients and upper/lower state 
IDs and divided into 10 files in 1000~cm$^{-1}$ ranges. An extract from a \CCC\ transition 
file is given in Table~\ref{t:trans}. We used the \CCC\ MARVEL energies of 
\citet{jt915} to replace the calculated values  where available. 
These entries are indicated in the states file with the label 'Ma'. 
The coverage of the MARVELised data is illustrated in Fig.~\ref{figMARVEL}, 
where we show a room temperature ($T=296$~K) spectrum of \CCC\ and the MARVELised 
lines indicated with the red circles. At this temperature, there are 16~178 MARVELised 
transitions which provide the experimental accuracy as based on the the experimental 
uncertainties of the MARVEL dataset by \citet{jt915}.
These high accuracy transitions can be accessed via the new ExoMolHR web app \citet{jt962}.

\begin{table*}
\centering
\caption{\label{t:states} Extract from the \texttt{.states} file of the \name\ \CTHREE\ line list. }
{\setlength{\tabcolsep}{2.5pt}
\footnotesize\tt
\begin{tabular}{rrrrrrcrrrrrrrrrrcr}
\hline \hline
        $i$  &  $\tilde{E}$/\cmi   &  $g$  &  $J$  & unc./\cmi & \multicolumn{1}{c}{$\tau$ / $s^{-1}$}  & $\Gamma_{\rm tot}$ & $n_1$ & $n_2^{\rm lin}$ & $l_2$ & $n_3$ & $\Gamma_{\rm vib}$& $C_i$ &  $v_1^{\rm T}$ & $v_2^{\rm T}$ & $v_3^{\rm T}$ & Ca/Ma  & \multicolumn{1}{c}{$\tilde{E}_{\rm T}$/\cmi} \\
 \hline
  39668 &       18.082083 &   13 &    6 &    0.000234 &         NaN  &  A1  &     0 &    0 &    0 &    0 &   A1  &    1.0000 &    0 &    0 &    0 &   Ma   &    18.082390\\
  39669 &      150.215573 &   13 &    6 &    0.005745 &   5.8667E+01 &  A1  &     0 &    0 &    2 &    0 &   A1  &    0.9900 &    0 &    0 &    0 &   Ma   &   150.216293\\
  39670 &      151.792684 &   13 &    6 &    0.001821 &   7.0628E+01 &  A1  &     0 &    0 &    0 &    1 &   A1  &   -0.9900 &    0 &    0 &    1 &   Ma   &   151.801811\\
  39671 &      301.417834 &   13 &    6 &    0.009863 &   2.4206E+01 &  A1  &     0 &    0 &    4 &    0 &   A1  &   -1.0000 &    0 &    0 &    0 &   Ma   &   301.414477\\
  39672 &      304.941414 &   13 &    6 &    0.005745 &   2.8200E+01 &  A1  &     0 &    0 &    2 &    1 &   A1  &   -0.9600 &    0 &    0 &    1 &   Ma   &   304.932485\\
  39673 &      306.313661 &   13 &    6 &    0.005745 &   3.0356E+01 &  A1  &     0 &    0 &    0 &    2 &   A1  &   -0.9600 &    0 &    0 &    2 &   Ma   &   306.289813\\
  39674 &      468.895204 &   13 &    6 &    0.004200 &   1.3788E+01 &  A1  &     0 &    0 &    6 &    0 &   A1  &    1.0000 &    0 &    0 &    0 &   Ca   &   468.895204\\
  39675 &      474.279718 &   13 &    6 &    0.014200 &   1.5230E+01 &  A1  &     0 &    0 &    4 &    1 &   A1  &   -1.0000 &    0 &    0 &    1 &   Ca   &   474.279718\\
  39676 &      477.142161 &   13 &    6 &    0.024200 &   1.6970E+01 &  A1  &     0 &    0 &    2 &    2 &   A1  &    0.9200 &    0 &    0 &    2 &   Ca   &   477.142161\\
\hline\hline
\end{tabular}}
\mbox{}\\

\mbox{}\\

{\flushleft
\begin{tabular}{ll}
\hline 
\noindent
$i$:&   State counting number.     \\
$\tilde{E}$:& State energy in \cmi. \\
$g_{\rm tot}$:& Total state degeneracy.\\
$J$:& Total angular momentum.            \\
unc.:& Uncertainty \cmi.     \\
$\tau$:& Life time in s.     \\
$\Gamma$:&   Total symmetry index in $C_{2v}$(M).\\
$n_1$:& Normal mode stretching N-N quantum number. \\
$n_2^{\rm lin}$:& Normal mode bending quantum number. \\
$l_2$:& Normal mode vibrational angular momentum quantum number. \\
$n_3$:& Normal mode stretching N-O quantum number. \\
$\Gamma_{\rm vib}$:&   Vibrational  symmetry index in $C_{2v}$(M).  \\
$C_i$:& Coefficient with the largest contribution to the $(J=0)$ contracted set; $C_i\equiv 1$ for $J=0$. \\
$v_1^{\rm T}$:&   TROVE stretching vibrational quantum number.\\
$v_2^{\rm T}$:&   TROVE\ stretching vibrational quantum number.\\
$v_3^{\rm T}$:&   TROVE bending vibrational quantum number.\\
Label:&  ``{\tt Ma}'' for MARVEL, ``{\tt Ca}'' for calculated.  \\
Calc:&  Original \textsc{TROVE} calculated state energy (in cm$^{-1}$).\\
\hline 
\end{tabular}
}
\end{table*}

\begin{table}
\centering
\caption{Extract from the transitions file for the \name\ line list for \CTHREE.  }
{\tt
\begin{tabular}{rrr}
\hline\hline
$f$ & $i$ & $A_{fi}$\\
\hline
      286793   &     299447  &2.6382e-16  \\
     1115453   &    1124869  &5.5991e-16  \\
     1144929   &    1134888  &7.8148e-16  \\
     1200446   &    1172433  &1.2177e-15  \\
     1329282   &    1320088  &2.4308e-16  \\
      879962   &     869764  &2.4688e-16  \\
     1464383   &    1455885  &7.3244e-16  \\
      616080   &     582311  &1.8432e-15  \\
      442578   &     430254  &2.8786e-16  \\
\hline\hline
\end{tabular}
}
\label{t:trans}
\mbox{}\\
{$f$}: Upper state counting number.  \\
{$i$}: Lower state counting number. \\
$A_{fi}$: Einstein~A-coefficient in s$^{-1}$.\\
\end{table}

\vskip 0.5cm

\section{Cross-Section, Partition Function and Opacity Estimates}

\citet{2018A+A...614A.131Y} provided a general utility program (ExoCross) for processing transition and states files generated by TROVE and other programs.  ExoCross was used to calculate cross-sections and partition functions based on transitions and corresponding ro-vibrational states calculated with TROVE
for the $\tilde{X}{\,}^{1}{\Sigma_{\rm g}}^{+}$ electronic ground-state of \CTHREE. 

\begin{figure}
\centering
\includegraphics[width=0.95\columnwidth]{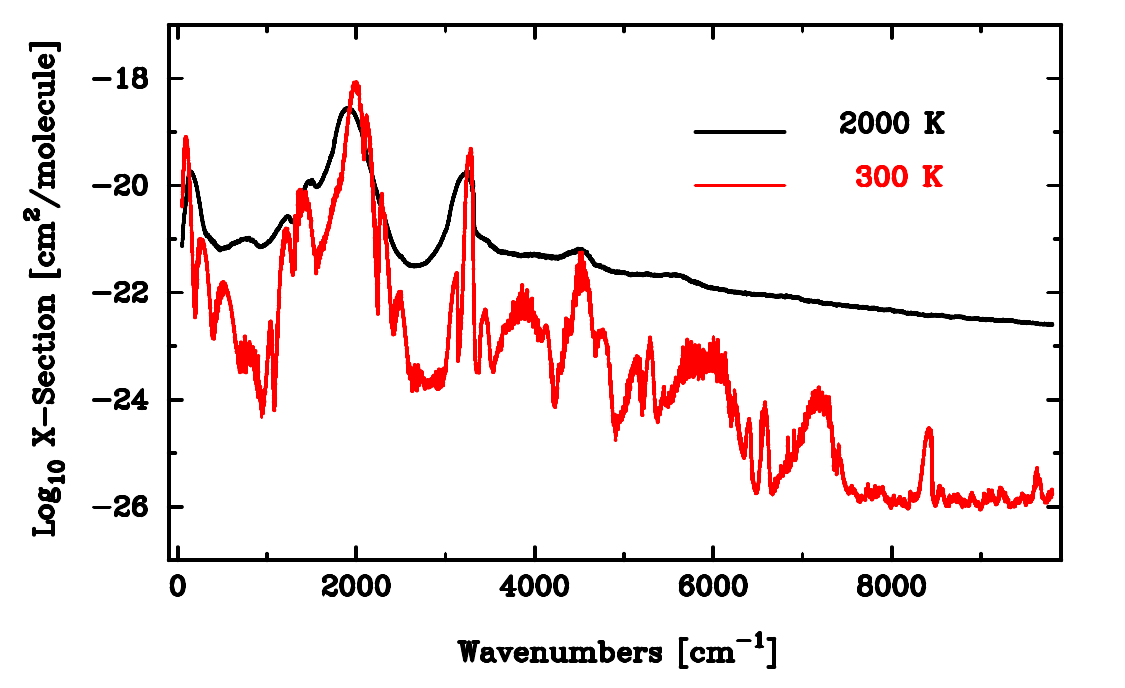}
\caption{Computed cross-sections for the
$\tilde{X}{\,}^{1}{\Sigma_{\rm g}}^{+}$ electronic ground-state of
\CCC\ at $300\,$~K and $2000\,$~K}
\label{figcccsec}
\end{figure}

\begin{figure}
\centering
\includegraphics[width=0.95\columnwidth]{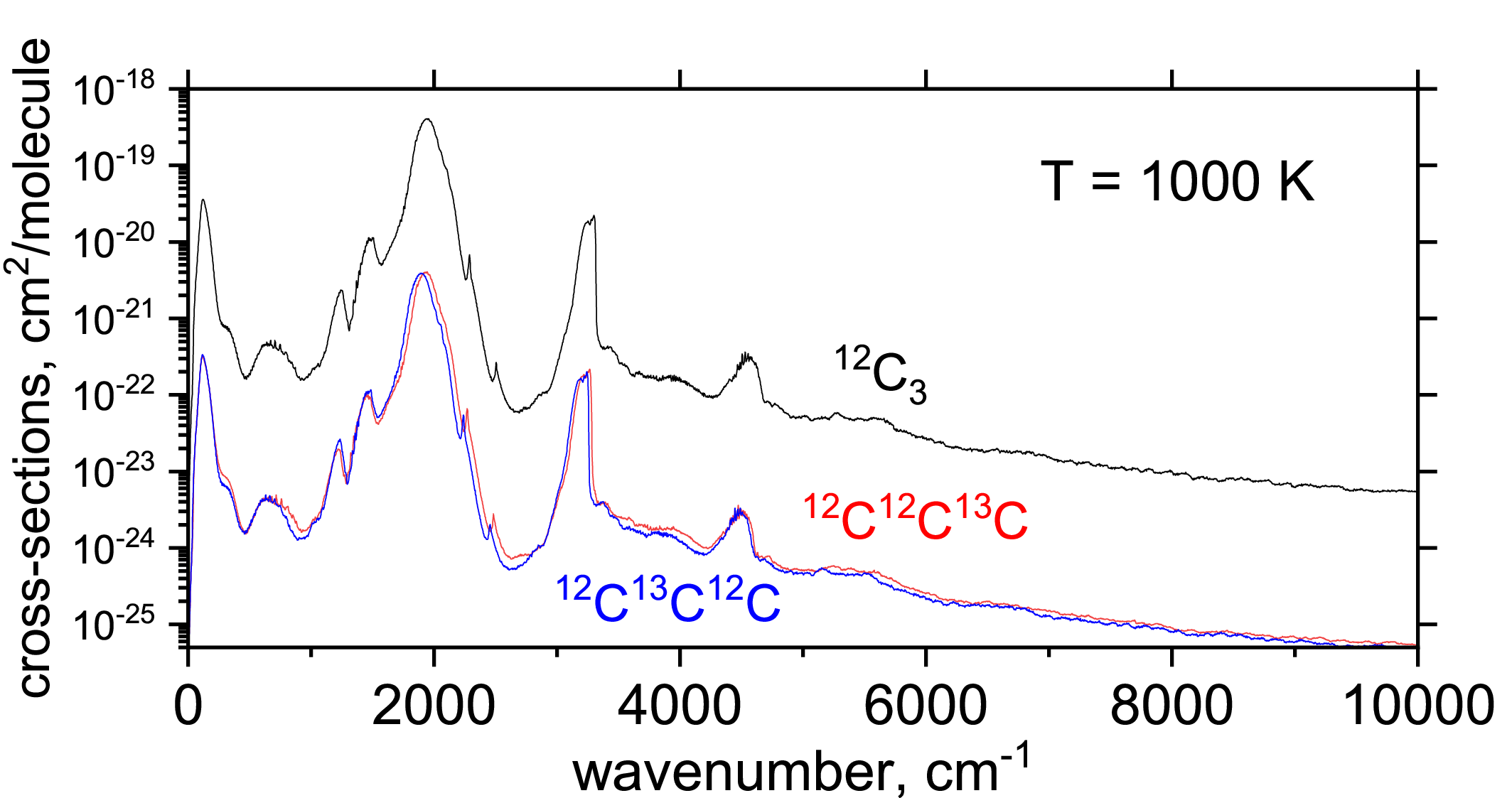}
\includegraphics[width=0.95\columnwidth]{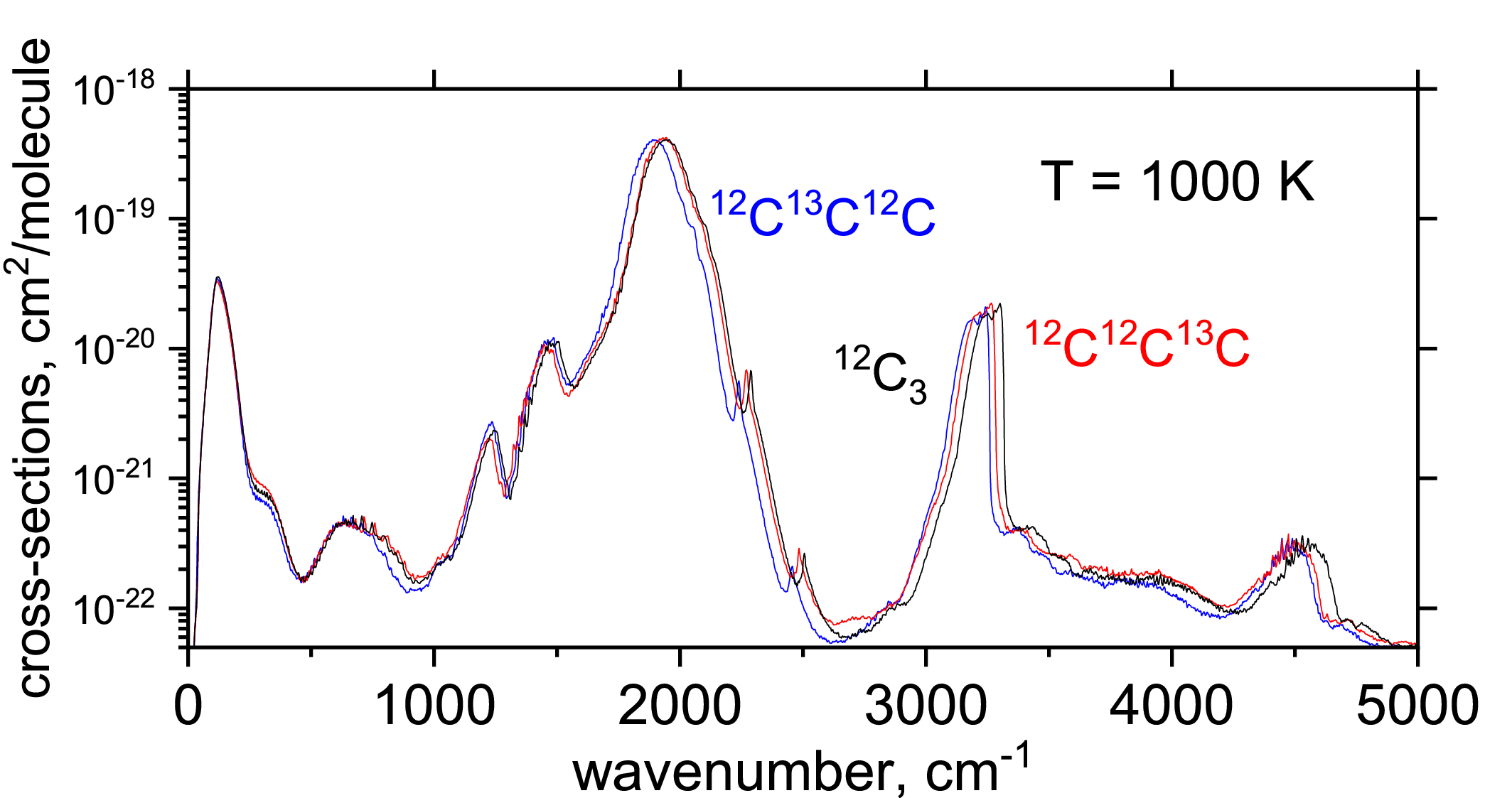}
\caption{Computed cross-sections for the
$\tilde{X}{\,}^{1}{\Sigma_{\rm g}}^{+}$ electronic ground-state of
\XIICTHREE, \CSC\ and \CCS\ at $1000\,$~K: scaled by their natural abundances (top display) to show their relative importance  and not-scaled (bottom display) to show the relative displacements.}
\label{xsecs}
\end{figure}

Figure~\ref{figcccsec} shows derived cross-sections,
at $300\,$~K (red curve) and $2000\,$~K (black curve),
for the \CCC\ electronic ground state 
sampled at 1~\cmi\ intervals and
convolved with a Gaussian of half-width at
half-maximum (HWHM) of 1~\cmi.  Of particular note is the peak near
2000~\cmi which was understood to correspond with the laboratory
measurement by \citet[their figure 2]{1975ApJ...202..839T}.
The weaker band at 3300~\cmi\ can also be identified.

\begin{figure}
\centering
\includegraphics[width=0.95\columnwidth]{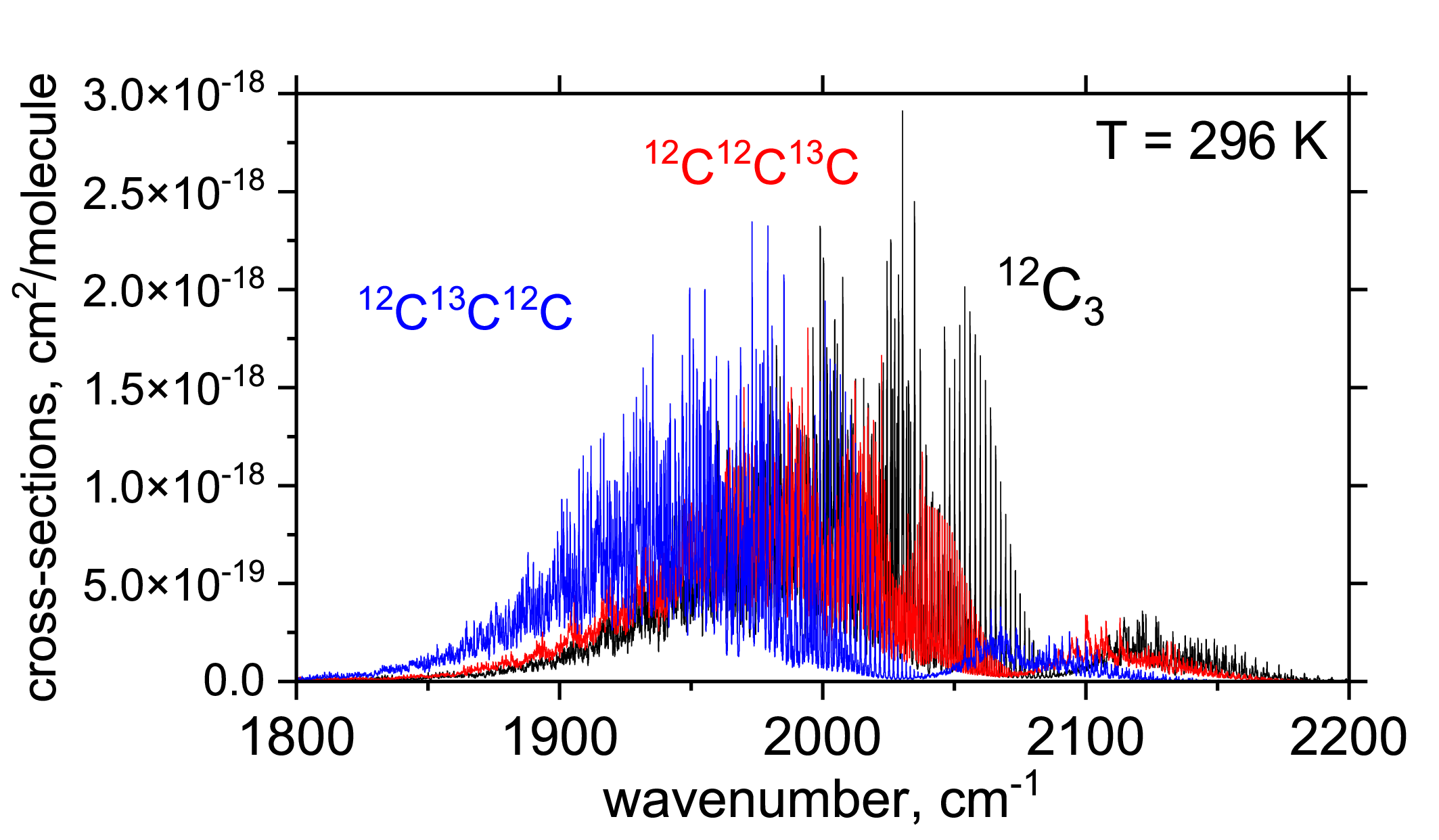}
\caption{Computed cross-sections for \CCC\ (black), \CCS\ (red)
and \CSC\ (blue) 2000 \cmi\ bands at a temperature of 296~K shown
as ``stick-spectra''.}
\label{T2000}
\end{figure}

\begin{figure}
\centering
\includegraphics[width=0.95\columnwidth]{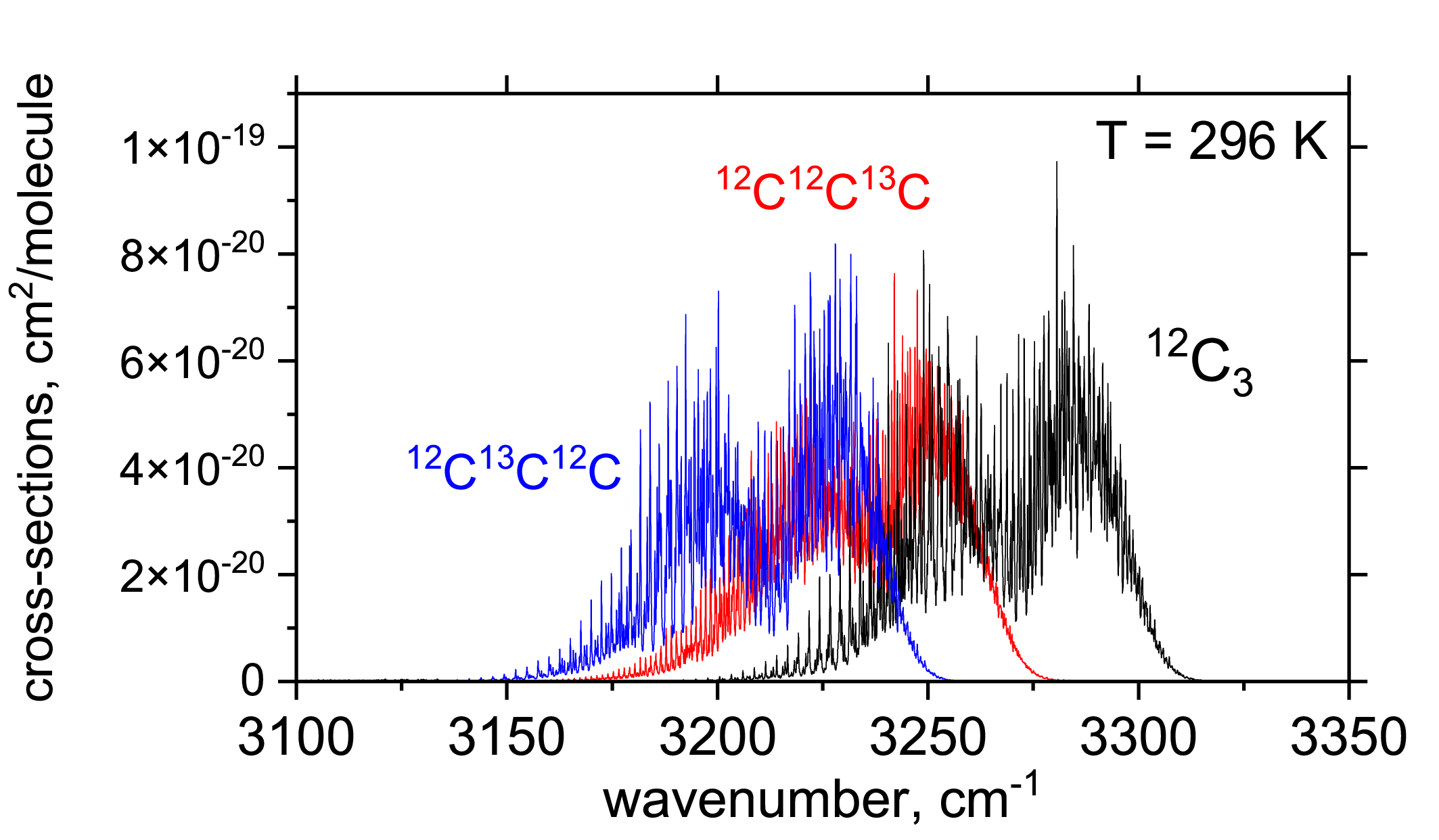}
\caption{Computed cross-sections for \CCC\ (black), \CCS\ (red)
and \CSC\ (blue) 3300 \cmi\ bands at a temperature of 296~K shown
as ``stick-spectra''}
\label{T3300}
\end{figure}

Cross-sections for all three isotopologues were similarly
calculated for electronic ground states at $1000\,$~K,
sampled at 1~\cmi\ intervals and convolved
with Gaussians of HWHM = 1~\cmi\ and  plotted in Figure~\ref{xsecs}. The top display, where  cross-sections are scaled by relative \XIICTHREE\ and \XIIICTHREE\
solar abundances, illustrates the relative importance of different isotopologues of C$_3$, while the bottom display, where the un-scaled cross-sections are overlaid, illustrates the isotope frequency shift between \CCC, \CSC\ and \CCS, which is discernable and can be seen even at low resolution in Figure~\ref{xsecs}. The consequences for the computed transition frequencies when a $^{13}{\mathrm C}$ replaces a $^{12}{\mathrm C}$ atom were more obvious in Figure~\ref{T2000} and Figure~\ref{T3300} where the 2040~\cmi\ and 3300~\cmi\ bands for all three isotopologues are plotted, showing differences of about 50~\cm. 


Temperature-dependent partition functions were obtained as Boltzmann sums over all states in a line list:
$$
Q(T)  = \sum_{i} g^{\rm (ns)} (2 J_i +1 ) e^{-c_2 \tilde{E}_i/T},
$$
where $\tilde{E}_i$ is the energy term value for state $i$ (\cm) calculated with TROVE for the $\tilde{X}{\,}^{1}{\Sigma_{\rm g}}^{+}$ electronic ground-state of \CTHREE, $J_i$ is the corresponding total angular momentum, $k$ is the second radiation constant (cm/K), $g^{\rm (ns)}$ is the state dependent nuclear spin degeneracy  and $T$ is the temperature (K). Since our partition functions only include states from the ground electronic states and therefore are not complete, especially at high temperatures, they 
thus represent lower limits.

\begin{figure}
\centering
\includegraphics[width=0.95\columnwidth]{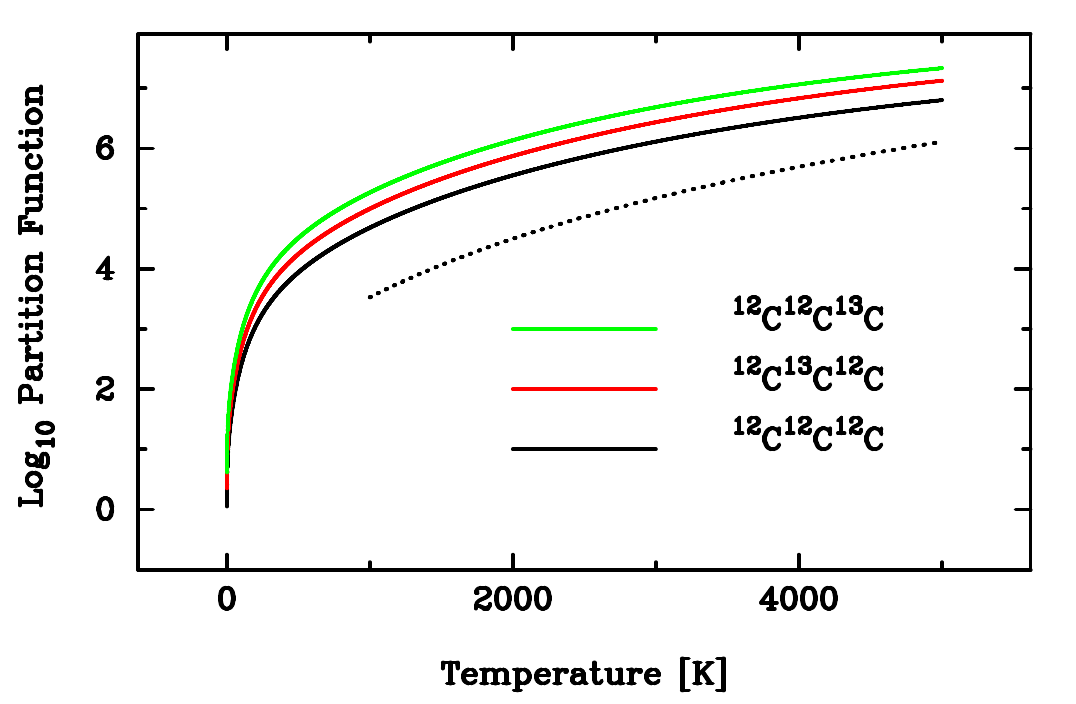}
\caption{Isotopologue partition function comparisons with a
\CTHREE\ estimate (dotted line) based on
\citeauthor{1981ApJS...45..621I}'s \citeyearpar{1981ApJS...45..621I}
polynomial coefficients}
\label{figpf}
\end{figure}

Note that ExoMol uses the HITRAN convention for partition functions which include the full nuclear spin degeneracy. The partition functions $Q(T)$ for \CCC, \CSC\ and \CCS\   were computed using temperatures in the range 1 to 5000~K in 1~K steps using ExoCross, which calculates
a finite sum over all calculated states following (for example) the approach which \citet{1995ApJ...454L.169N} adopt.  The resulting dependence of partition function with temperature was plotted in Figure~\ref{figpf}.  Given energy maxima and the $J$-range selected (Table~\ref{trveio}), partition functions plotted in Figure~\ref{figpf}
were fully converged. \citeauthor{1981ApJS...45..621I}'s \citeyearpar{1981ApJS...45..621I} polynomial coefficients lead to a \CCC\ partition function lower than our estimate by roughly one order of magnitude; this was anticipated as his polynomial coefficients are based on studies by \citet{1959UCRL8675......P} and \citet{1963NASSP3001.....M}, predating pioneering work by \citet{1965ApJ...142...45G} who demonstrated the importance of the contribution from the low frequency bending mode.  Since $^{12}$C has zero nuclear spin, there is no difference for \CCC, but for the $^{13}C$-containing isotopologues considered here, our partition functions will be a factor of two larger than standard astrophysical ones which neglect the nuclear spin factor.

Apart from the states, transition and partition function files, ExoMol line list 
compilations contain opacities and line broadening parameters. Opacities in the form of
cross-sections and k-tables \citep{1991JGR....96.9027L}
were calculated following the procedure \citet{2021A+A...646A..21C} describe,
adopting their temperature and pressure grid.
Molecular opacities for C$_3$ are provided in four files, each formatted for 
the exoplanet atmospheric retrieval code identified in the file name: 
ARCiS~\citep{ARCiS}, TauREx~\citep{TauRex3}, NEMESIS~\citep{NEMESIS} and 
petitRADTRANS~\citep{19MoWaBo.petitRADTRANS}.

As \citet[their figure 1]{1984A+A...132..236B} show, an accurate
determination of the microturbulent velocity is required for a solar
abundance determination based on several absorption lines, 
using precision oscillator strengths and a one-dimensional static model
stellar atmosphere in which radiation transfer occurs only along an 
observer's line-of-sight.
\citet{2000A+A...359..729A} use more realistic dynamical model 
atmospheres, in which three-dimensional radiation transfer is coupled
with the convective velocity field; they show that a
microturbulent velocity is an artefact of the static one-dimensional 
model approximation and not needed.   Retrieval codes for which
we have provided \CTHREE\ opacities make no use of microturbulent
velocity and we have accordingly ignored it in our opacity table
calculations.

\section{Comparison with Experiment}

\begin{figure*}
\centering
\includegraphics[width=1.95\columnwidth]{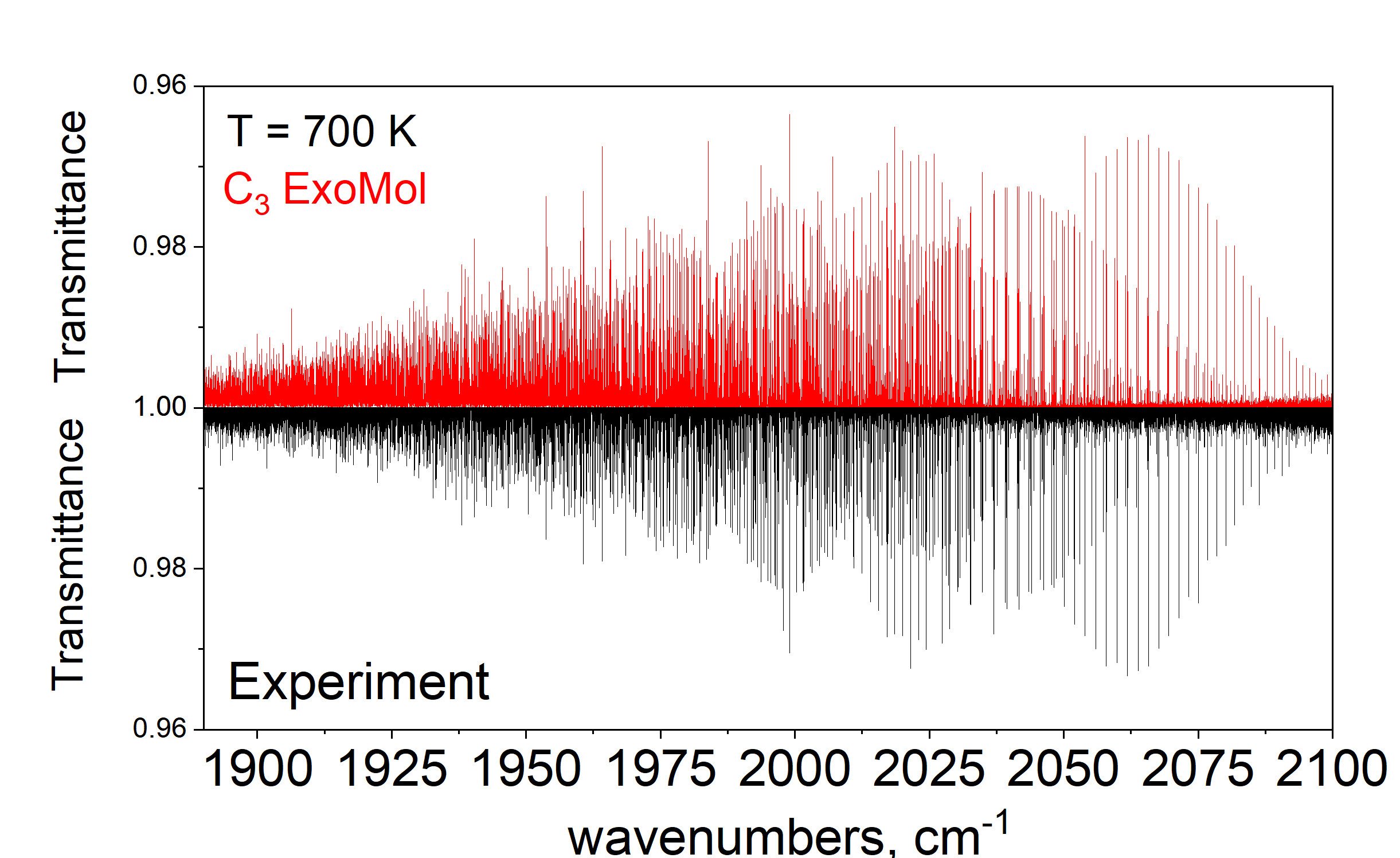}
\caption{The \CTHREE\ experimental  spectrum due to \citet{2023JMoSp.39111734M} with the transmittance 
compared to the transmittance computed (see text) for their experimental conditions using our \XIICTHREE\ line list and plotted in the lower panel. The irregular spikes in  both spectra correspond to local accidental coincidences of line positions. These responses are very sensitive to the accuracy of calculations as well as experimental resolutions and are very difficult to reproduce, which explains their appearance at different places. 
}
\label{figcomp}
\end{figure*}

\citet[their figure 5]{2023JMoSp.39111734M} present a Fourier
transform infrared spectrum of \CTHREE\ in absorption from 1900 to
2100~\cmi, deducing a 700~K rotational temperature and a Gaussian
line-shape of width 0.005~\cmi\ as their experimental conditions.
Computed \CTHREE\ transitions and state energies were used with ExoCross to synthesise the \citeauthor{2023JMoSp.39111734M} spectrum, assuming the
same  temperature and Gaussian width.  The comparison is shown in
Figure~\ref{figcomp}: black is used for the
\citeauthor{2023JMoSp.39111734M} spectrum and red for the spectrum
synthesised using \CTHREE\ transitions and energies reported in the
present paper.  The wavenumber range plotted in Figure~\ref{figcomp}
corresponds to the $v_3 \simeq 2000$~\cmi\ broad absorption feature
plotted in Figs.~\ref{figcccsec} and \ref{xsecs}.

A high resolution inspection of individual lines plotted in
Figure~\ref{figcomp} showed a maximum difference between observed and calculated line-positions of 0.03~\cmi, as expected given the formal least squares errors obtained using MARVEL energies for PES refinement described above.  In most cases, calculated and observed line-positions agreed to better than 0.01~\cmi\ as indicated by the state energy
comparison presented in Table~\ref{tabstates}.
Our calculated line-intensities are also in good agreement with the  observed relative values,
as may be seen in Figure~\ref{figcomp}, showing the band structure to
have been correctly calculated. Our values can be used to place observations on an absolute scale.

\setlength{\tabcolsep}{1pt}
\begin{table}
\begin{center}
\caption{Comparisons of theoretical vibrational term values of \CTHREE\ isotopologues with experimentally derived values from the literature. }
\label{kriegmolp}
\begin{tabular}{ccd{5.7}d{5.7}d{5.7}}
\hline
             &             &              &                  &	           \\
Isotopologue & Vibrational & \multicolumn{1}{c}{\citeauthor{2013JPCA..117.3332K}}
                           & \multicolumn{1}{c}{Other}
                           & \multicolumn{1}{c}{TROVE}                     \\
             &    State    & \multicolumn{1}{c}{(\cmi)}
                           & \multicolumn{1}{c}{(\cmi)}
                           & \multicolumn{1}{c}{(\cmi)}                    \\
             &             &              &                  &             \\
\hline
             &             &              &                  &             \\
\CCC         &  \OIIO      &   63.4165946 &   63.4165889     &   63.415212 \\
             &  \OOOI      & 2040.019278  &                  & 2040.017256 \\
             &  \OIII      & 2078.500541  &                  & 2078.48975  \\
             &  \IOOI      & 3260.127048  &                  & 3260.974851 \\
             &  \IIII      & 3330.508589  &                  & 3330.503579 \\
             &             &              &                  &             \\
\CCS         &  \OOOI      & 2027.20779   &  2027.2078       & 2027.201829 \\
             &  \IOOI      & 3224.7509    &                  & 3224.733608 \\
             &             &              &                  &             \\
\CSC         &  \IOOI      & 3205.59319   &                  & 3205.561181 \\
             &             &              &                  &             \\
\hline
\end{tabular}
\end{center}
\end{table}

Best fit molecular vibrational term values (band centres)  by
\citet[their tables 3, 5 and 7]{2013JPCA..117.3332K} for \CCC, \CSC\ and \CCS\ 
are listed in Table~\ref{kriegmolp} (Column 3) for comparison with our ($J=0$) energies of \CTHREE. 
The \CCC\ \OIIO\ energy term value in Column 4 was taken from
\citet[their table 2]{2003ZNatA..58..129G} and the \CCS\ \OOOI\ value in the
same column was a refit to the measurements of \citet{1993JChPh..98.7757M}  by
\citet[their table 7]{2013JPCA..117.3332K}.  With the exception of the
\CCC\ \IOOI\ entry where the difference was larger than 0.8 \cmi, satisfactory
agreement is obtained. This is especially reassuring considering that our model (PES) was optimised for the main isotopologue only. 

\setlength{\tabcolsep}{4pt}
\begin{table}
\begin{center}
\caption{Comparison of Observed \citep{2016JChPh.145w4302B}
and Calculated $(0,1^1,0)-(0,0^0,0)$ Transition Frequencies}
\label{breiercomp}
\begin{tabular}{ccrccd{5.4}}
\hline
             &        &    &             &             &           \\
Isotopologue & Branch &  J &   Observed  &  Calculated & \multicolumn{1}{c}{\OMC}   \\
             &        &    &    (\cmi)   &   (\cmi)    & \multicolumn{1}{c}{(\cmi)} \\
             &        &    &             &             &           \\
\hline
             &        &    &             &             &           \\
   \CCC      &    P   &  2 &    61.2698  &  61.2698    &  -0.0000  \\
             &        &  4 &    59.6376  &  59.6376    &  -0.0000  \\
             &        &  6 &    58.0776  &  58.0775    &   0.0001  \\
             &        &  8 &    56.5900  &  56.5900    &   0.0000  \\
             &        & 10 &    55.1744  &  55.1744    &   0.0000  \\
             &        & 12 &    53.8307  &  53.8307    &   0.0000  \\
             &        & 14 &    52.5584  &  52.5582    &   0.0002  \\
             &        & 16 &    51.3567  &  51.3566    &   0.0000  \\
             &        &    &             &             &           \\
   \CCC      &    Q   &  2 &    63.0622  &  63.0622    &   0.0000  \\
             &        &  4 &    63.2673  &  63.2673    &   0.0000  \\
             &        &  6 &    63.5886  &  63.5886    &  -0.0000  \\
             &        &  8 &    64.0247  &  64.0247    &   0.0000  \\
             &        & 10 &    64.5737  &  64.5737    &   0.0000  \\
             &        & 12 &    65.2334  &  65.2334    &  -0.0000  \\
             &        & 14 &    66.0012  &  66.0012    &   0.0000  \\
             &        & 16 &    66.8741  &  66.8741    &  -0.0000  \\
             &        &    &             &             &           \\
   \CCC      &    R   &  0 &    63.8533  &  63.8533    &  -0.0000  \\
             &        &  2 &    65.6653  &  65.6653    &   0.0000  \\
             &        &  4 &    67.5485  &  67.5484    &   0.0000  \\
             &        &  6 &    69.5023  &  69.5023    &  -0.0000  \\
             &        &  8 &    71.5262  &  71.5262    &  -0.0000  \\
             &        & 10 &    73.6193  &  73.6193    &  -0.0000  \\
             &        & 12 &    75.7806  &  75.7806    &  -0.0001  \\
             &        & 14 &    78.0089  &  78.0090    &  -0.0000  \\
             &        & 16 &    80.3032  &  80.3027    &   0.0005  \\
             &        &    &             &             &           \\
   \CSC      &    P   &  2 &    58.9045  &  58.8754    &   0.0291  \\
             &        &  4 &    57.2690  &  57.2397    &   0.0293  \\
             &        &  6 &    55.7018  &  55.6727    &   0.0291  \\
             &        &  8 &    54.2020  &  54.1744    &   0.0276  \\
             &        & 10 &    52.7683  &  52.7448    &   0.0235  \\
             &        & 12 &    51.3987  &  51.3834    &   0.0153  \\
             &        &    &             &             &           \\
   \CSC      &    Q   &  2 &    60.6952  &  60.6663    &   0.0289  \\
             &        &  4 &    60.8970  &  60.8681    &   0.0290  \\
             &        &  6 &    61.2132  &  61.1841    &   0.0291  \\
             &        &  8 &    61.6423  &  61.6131    &   0.0292  \\
             &        & 10 &    62.1824  &  62.1532    &   0.0293  \\
             &        & 12 &    62.8312  &  62.8020    &   0.0292  \\
             &        &    &             &             &           \\
   \CSC      &    R   &  0 &    61.4863  &  61.4573    &   0.0289  \\
             &        &  2 &    63.2928  &  63.2638    &   0.0290  \\
             &        &  4 &    65.1667  &  65.1380    &   0.0287  \\
             &        &  6 &    67.1064  &  67.0792    &   0.0272  \\
             &        &  8 &    69.1102  &  69.0870    &   0.0232  \\
             &        & 10 &    71.1755  &  71.1603    &   0.0152  \\
             &        & 12 &    73.2996  &  73.2986    &   0.0010  \\
             &        &    &             &             &           \\
   \CCS      &    P   &  2 &    61.0766  &  61.0770    &  -0.0004  \\
             &        &  3 &    60.2841  &  60.2845    &  -0.0003  \\
             &        &  4 &    59.5091  &  59.5094    &  -0.0003  \\
             &        &  5 &    58.7516  &  58.7518    &  -0.0001  \\
             &        &  6 &    58.0117  &  58.0116    &   0.0001  \\
             &        &  7 &    57.2893  &  57.2890    &   0.0003  \\
             &        &  8 &    56.5844  &  56.5838    &   0.0007  \\
             &        &  9 &    55.8971  &  55.8960    &   0.0011  \\
             &        & 10 &    55.2274  &  55.2257    &   0.0017  \\
             &        & 11 &    54.5751  &  54.5727    &   0.0024  \\
             &        & 12 &    53.9403  &  53.9370    &   0.0034  \\
\hline
\end{tabular}
\end{center}
\end{table}

\begin{table}
\begin{center}
\contcaption{}
\begin{tabular}{ccrccd{5.4}}
\hline
             &        &    &             &             &           \\
             &        &    &             &             &           \\
Isotopologue & Branch &  J &   Observed  &  Calculated & \multicolumn{1}{c}{\OMC}   \\
             &        &    &    (\cmi)   &   (\cmi)    & \multicolumn{1}{c}{(\cmi)} \\
             &        &    &             &             &           \\
\hline
             &        &    &             &             &           \\
   \CCS      &    Q   &  1 &    62.7421  &  62.7425    &  -0.0004  \\
             &        &  2 &    62.7981  &  62.7984    &  -0.0004  \\
             &        &  3 &    62.8819  &  62.8823    &  -0.0003  \\
             &        &  4 &    62.9936  &  62.9939    &  -0.0003  \\
             &        &  5 &    63.1331  &  63.1333    &  -0.0002  \\
             &        &  6 &    63.3000  &  63.3001    &  -0.0001  \\
             &        &  7 &    63.4944  &  63.4944    &  -0.0000  \\
             &        &  8 &    63.7159  &  63.7159    &   0.0001  \\
             &        &  9 &    63.9644  &  63.9643    &   0.0002  \\
             &        & 10 &    64.2396  &  64.2394    &   0.0002  \\
             &        & 11 &    64.5411  &  64.5410    &   0.0002  \\
             &        & 12 &    64.8687  &  64.8687    &   0.0001  \\
             &        &    &             &             &           \\
   \CCS      &    R   &  0 &    63.5590  &  63.5594    &  -0.0004  \\
             &        &  1 &    64.4212  &  64.4216    &  -0.0004  \\
             &        &  2 &    65.3008  &  65.3011    &  -0.0003  \\
             &        &  3 &    66.1977  &  66.1978    &  -0.0001  \\
             &        &  4 &    67.1118  &  67.1117    &   0.0001  \\
             &        &  5 &    68.0431  &  68.0428    &   0.0004  \\
             &        &  6 &    68.9916  &  68.9908    &   0.0008  \\
             &        &  7 &    69.9571  &  69.9558    &   0.0013  \\
             &        &  8 &    70.9396  &  70.9376    &   0.0020  \\
             &        &  9 &    71.9390  &  71.9362    &   0.0028  \\
             &        & 10 &    72.9553  &  72.9514    &   0.0039  \\
             &        & 11 &    73.9882  &  73.9831    &   0.0052  \\
             &        & 12 &    75.0379  &  75.0312    &   0.0067  \\
\hline
\end{tabular}
\end{center}
\end{table}

\begin{table}
\begin{center}
\caption{Comparison of Observed \citep{2013JPCA..117.3332K}
and Calculated \IOOI -- \OOOO Transition Frequencies}
\label{kriegcomp}
\begin{tabular}{ccrccd{5.4}}
\hline
             &        &    &             &             &           \\
Isotopologue & Branch &  J &   Observed  &  Calculated & \multicolumn{1}{c}{\OMC}   \\
             &        &    &    (\cmi)   &   (\cmi)    & \multicolumn{1}{c}{(\cmi)} \\
             &        &    &             &             &           \\
\hline
             &        &    &             &             &           \\
\CCC         &  P     &  2 &   3258.3915 &  3258.3914  &   +0.0001 \\
             &        &  4 &   3256.6061 &  3256.6061  &   -0.0000 \\
             &        &  6 &   3254.7701 &  3254.7704  &   -0.0003 \\
             &        &  8 &   3252.8857 &  3252.8858  &   -0.0001 \\
             &        & 10 &   3250.9513 &  3250.9514  &   -0.0001 \\
             &        & 12 &   3248.9690 &  3248.9702  &   -0.0012 \\
             &        & 14 &   3246.9399 &  3246.9396  &   +0.0003 \\
             &        & 16 &   3244.8634 &  3244.8637  &   -0.0003 \\
             &        & 18 &   3242.7418 &  3242.7419  &   -0.0001 \\
             &        & 20 &   3240.5760 &  3240.5776  &   -0.0016 \\
             &        & 22 &   3238.3673 &  3238.3673  &   -0.0000 \\
             &        & 24 &   3236.1143 &  3236.1143  &   +0.0000 \\
             &        & 26 &   3233.8189 &  3233.8189  &   +0.0000 \\
             &        &    &             &             &           \\
\CCC         &  R     &  0 &   3260.9744 &  3260.9749  &   -0.0005 \\
             &        &  2 &   3262.6340 &  3262.6338  &   +0.0002 \\
             &        &  4 &   3264.2415 &  3264.2414  &   +0.0001 \\
             &        &  6 &   3265.7983 &  3265.7982  &   +0.0001 \\
             &        &  8 &   3267.3032 &  3267.3032  &   -0.0000 \\
             &        & 10 &   3268.7595 &  3268.7587  &   +0.0008 \\
             &        & 12 &   3270.1620 &  3270.1620  &   -0.0000 \\
             &        & 14 &   3271.5163 &  3271.5160  &   +0.0003 \\
             &        & 16 &   3272.8208 &  3272.8207  &   +0.0001 \\
             &        & 18 &   3274.0776 &  3274.0775  &   +0.0001 \\
             &        & 20 &   3275.2857 &  3275.2851  &   +0.0006 \\
             &        & 22 &   3276.4472 &  3276.4454  &   +0.0018 \\
             &        & 24 &   3277.5579 &  3277.5569  &   +0.0010 \\
             &        & 26 &   3278.6059 &  3278.6059  &   +0.0000 \\
             &        &    &             &             &           \\
\CSC         &  P     &  2 &   3203.8601 &  3203.8279  &   +0.0322 \\
             &        &  4 &   3202.0786 &  3202.0467  &   +0.0319 \\
             &        &  6 &   3200.2508 &  3200.2182  &   +0.0326 \\
             &        &  8 &   3198.3737 &  3198.3429  &   +0.0308 \\
             &        & 10 &   3196.4519 &  3196.4212  &   +0.0307 \\
             &        & 12 &   3194.4845 &  3194.4541  &   +0.0304 \\
             &        &    &             &             &           \\
             &  R     &  0 &   3206.4418 &  3206.4098  &   +0.0320 \\
             &        &  2 &   3208.1020 &  3208.0709  &   +0.0311 \\
             &        &  4 &   3209.7142 &  3209.6835  &   +0.0307 \\
             &        &  6 &   3211.2776 &  3211.2477  &   +0.0299 \\
             &        &  8 &   3212.7938 &  3212.7634  &   +0.0304 \\
             &        & 10 &   3214.2615 &  3214.2310  &   +0.0305 \\
             &        & 12 &   3215.6799 &  3215.6506  &   +0.0293 \\
             &        &    &             &             &           \\
\CCS         &  P     &  1 &   3223.9235 &  3223.9062  &   +0.0173 \\
             &        &  2 &   3223.0837 &  3223.0665  &   +0.0172 \\
             &        &  3 &   3222.2317 &  3222.2146  &   +0.0171 \\
             &        &  4 &   3221.3654 &  3221.3506  &   +0.0148 \\
             &        &  5 &   3220.4872 &  3220.4744  &   +0.0128 \\
             &        &  6 &   3219.5997 &  3219.5863  &   +0.0134 \\
             &        &  7 &   3218.7039 &  3218.6862  &   +0.0177 \\
             &        &  8 &   3217.7917 &  3217.7742  &   +0.0175 \\
             &        &  9 &   3216.8658 &  3216.8505  &   +0.0153 \\
             &        & 10 &   3215.9316 &  3215.9150  &   +0.0166 \\
             &        & 11 &   3214.9840 &  3214.9679  &   +0.0161 \\
             &        &    &             &             &           \\
\hline
\end{tabular}
\end{center}
\end{table}

\begin{table}
\begin{center}
\contcaption{ }
\begin{tabular}{ccrccd{5.4}}
\hline
             &        &    &             &             &           \\
             &        &    &             &             &           \\
Isotopologue & Branch &  J &   Observed  &  Calculated & \multicolumn{1}{c}{\OMC}   \\
             &        &    &    (\cmi)   &   (\cmi)    & \multicolumn{1}{c}{(\cmi)} \\
             &        &    &             &             &           \\
\hline
             &        &    &             &             &           \\
\CCS         &  R     &  0 &   3225.5661 &  3225.5488  &   +0.0173 \\
             &        &  1 &   3226.3700 &  3226.3517  &   +0.0183 \\
             &        &  2 &   3227.1603 &  3227.1422  &   +0.0181 \\
             &        &  3 &   3227.9361 &  3227.9205  &   +0.0156 \\
             &        &  4 &   3228.7025 &  3228.6864  &   +0.0161 \\
             &        &  5 &   3229.4566 &  3229.4400  &   +0.0166 \\
             &        &  6 &   3230.1982 &  3230.1812  &   +0.0170 \\
             &        &  7 &   3230.9261 &  3230.9102  &   +0.0159 \\
             &        &  8 &   3231.6438 &  3231.6269  &   +0.0169 \\
             &        &  9 &   3232.3485 &  3232.3314  &   +0.0171 \\
             &        & 10 &   3233.0410 &  3233.0237  &   +0.0173 \\
             &        & 11 &   3233.7216 &  3233.7039  &   +0.0177 \\
\hline
\end{tabular}
\end{center}
\end{table}

\citet[their table 3]{2016JChPh.145w4302B} use a terahertz-supersonic jet spectrometer,
in combination with a laser ablation source, to measure $\nu_2$ lowest
bending mode transition frequencies of linear \CTHREE\ and its
$^{13}{\rm C}$-substituted isotopologues.   Table~\ref{breiercomp}
compares the $\nu_2$ lowest order bending mode transition frequencies
for \CCC, \CSC\ and \CCS\ extracted from our line lists with the
\citeauthor{2016JChPh.145w4302B} measurements.  In the \CCC\ case,
calculated frequencies agree with experiment to 0.0005 \cmi\ in the
worst case ($J = 16$, R-Branch); this was understood to be a consequence
of MARVEL corrections having been applied and PES improvements that these
made possible.  The \CCC\ \IOOI\ frequency
\citeauthor{2013JPCA..117.3332K}
deduce by fitting molecular parameters was noted above as differing from
our states file entry by more than 0.8~\cmi; this was reduced to 0.0005~\cmi\
when the \citet[their table 3]{2016JChPh.145w4302B}
measurement was adopted, as shown in Table~\ref{breiercomp}.

Line lists for \CSC\ and \CCS\ have been prepared without MARVEL corrections
as, to the best of our knowledge, necessary experimental data do not currently
exist.  As a result, the Table~\ref{breiercomp} comparison with experiment
for $^{13}{\rm C}$-substituted isotopologues was less satisfactory.
We also noted that use of a PES refined for \CCC\
would have been non-optimum for \CSC\ and \CCS\ line list calculations.
While $\nu_2$ lowest bending mode transition frequency residuals for \CSC\
were found to be always less than 0.03~\cmi\, they were all positive and
therefore indicative of an underlying  systematic error.

\cite{2013JPCA..117.3332K} conduct infrared $(\sim 3\,{\mu}{\mathrm m})$
high resolution spectroscopy of \CTHREE, also using a supersonic jet
and a laser ablation source, in combination
with a continuous-wave parametric oscillator as a radiation source.
In Table~\ref{kriegcomp} we have compared our wavenumbers for
$(101) \leftarrow (000)$ transitions with those
\citet[their tables 1, 4 and 6]{2013JPCA..117.3332K} measure.
In the \CCC\ case, calculated frequencies agree with experiment to
0.0018 \cmi\ in the worst case ($J = 22$, R-Branch); this was again
understood to be a consequence of MARVEL corrections having been applied
and PES improvements that these made possible.

As with the \citet{2016JChPh.145w4302B} comparison, the Table~\ref{kriegcomp}
comparison with experiment for $^{13}{\rm C}$-substituted isotopologues was
less satisfactory.  The absence of negative residuals is again apparent
and indicative of an underlying  systematic error
in our \CSC\ and \CCS\ calculations.  Identifying the origin of the apparent
systematic error in our \CSC\ and \CCS\ line list calculations was regarded
as beyond the scope of the present study but an obvious first step would
be to test the validity of the PES used, especially as this had been carefully
refined to optimise \CCC\ calculations.

\section{Discussion}

\citet{1975ApJ...202..839T} showed experimentally that the
$v_3 \simeq 2000$~\cmi\ broad absorption feature completely dominates
the integrated \CTHREE\ cross-section at 3100~K, with transitions
calculated in the present paper showing that this extends downwards to
temperatures of 300~K.  The $v_3 \simeq 2000$~\cmi\ band was therefore
found to completely dominate the infrared \CTHREE\ opacity as
\citet{1973IAUS...52..517G} suggests.  As \CTHREE\ transitions
computed in the present paper reproduce the $v_3 \simeq 2000$~\cmi\
band \citet{2023JMoSp.39111734M} observe, it was clear that our
line lists could be used to calculate the \CTHREE\ contribution to
molecular opacity at temperatures less than ${\sim}5000$~K; this limit
is a consequence of the fact that our lower state energies were 
limited to states below 10000~\cmi.

\citet{2022ApJ...940..129M} provide low-temperature gas
opacities.  In the \CTHREE\ case, \citeauthor{2022ApJ...940..129M}
estimate this contribution using the line list by
\citet{1989ApJ...343..554J}.  Our \CCC\ line list is complete to
transition frequencies of 10000~\cmi\ or less and our associated states
file complete for electronic ground state ro-vibrational energies less
than 20000~\cmi; it is therefore an improvement on the earlier
line list by \citeauthor{1989ApJ...343..554J}, especially as our
calculated stretching frequencies are in very much better agreement with
experimental determinations.  An immediate application of our line lists
for \CTHREE\ and its $^{13}{\rm C}$-substituted isotopologues would be
to update \citeauthor{2022ApJ...940..129M} low-temperature gas
opacities. As part of this work, a similar update has been applied to the ExoMolOP Database \citep{2021A+A...646A..21C} which did not include a \CTHREE\ contribution.

Use of high-dispersion transit spectroscopy to detect and characterise
extra-solar planets, by \citet{2019A+A...625A.107G} for example,
requires line lists for individual molecules having accurate
line-positions and relative intensities.  Our \CCC\ line list was
understood to be suitable for high-dispersion transit spectroscopy at
wavelengths longer than $\sim 2.5\,{\mu}{\mathrm m}$
($\sim 4000\,$\cmi).  Making our \CCC\ line list
suitable for high-dispersion transit spectroscopy at shorter wavelengths
would be complicated by the presence of conical intersections as
\citet{1996RvMP...68..985Y}, \citet{2004ARPC...55..127W} and
\citet{2018ARPC...69..427S} review.  Briefly, in the Born-Oppenheimer
approximation for molecular dynamics, nuclei move on potential energy
surfaces corresponding to associated electronic states; these may
intersect at a point in nuclear coordinates with the topology of a
double cone.  While \citet{2018PCCP...2010319R} improve on the PES by
\citet{2015JChPh.143g4302R} and achieve an accurate representation in
the vicinity of the four conical intersections, they do not claim an
improvement for other geometries.  Further PES development work will
therefore be needed.

As \citet{1988Sci...241.1319H} note, the 4050 \AA\ line cannot
be used to detect \CTHREE\ in the interstellar medium and cool stars
through the lack of ultraviolet flux; they identify this molecule
in the circumstellar shell surrounding the obscured carbon star
IRC$+10216$ using observed line positions from which molecular constants
were deduced.  Detection of \CTHREE, where it exits, is important as
chemical models of the environment are then constrained and our \CCC\
line list should facilitate searches when infrared or radio data
are available.  A kinematic analysis using lines of molecular species,
identified and for which number densities are known through a reliable
chemical model, establishes star formation modes as
\citet{2019ApJ...871..134S} demonstrate.

As it is understood from stellar evolution calculations 
\citep{1989ApJS...69..911S} that $^{13}{\mathrm C}$ will be enriched 
in stars ascending the giant and asymptotic giant branches due to dredge-up,
and \citet{2020A+A...633A.120G} detect \CSC\ and \CCS\ in the
interstellar medium towards Sgr~B2(M), future work to improve
our line lists for these isotopologues would seem to be worthwhile.
More laboratory work to create experimental line lists for
\CSC\ and \CCS\ would enable the MARVEL procedure to be applied;
PES refinement should follow for both cases, allowing improved
line lists to be calculated.  

\citet{2024A+A...684A..70L} obtain \XIIXIII\ abundance ratios for
seventy-one field red giants and identify cases where 
\XIIXIII$\,\leqslant 10$.  The relative contribution to stellar interior 
opacity, and therefore to stellar structure, by isotopologues having 
one or more $^{12}{\mathrm C}$ substituted by a $^{13}{\mathrm C}$ atom, 
will become increasingly important as the \XIIXIII\ ratio decreases.
Additional line lists for the remaining isotopologues
(\SCS, \SSC\ and \SSS) could be computed should the need arise.

\section{Conclusions}

We have calculated new  \name\ line lists for the electronic ground
state of the \CTHREE\ molecule and its two $^{13}{\rm C}$-substituted
isotopologues, known to be important in astrophysics. We have made the
transition, states and partition function files available through the
ExoMol database.  Our line lists are applicable to studies of the
interstellar medium as well as cool star and planetary atmospheres.
In the \CCC\ case, the MARVEL procedure has provided experimental
accuracy in all comparisons; the other two line lists are not as
accurate, and improvements to be made in the future have been
identified.

\section*{Acknowledgements}

The authors are indebted to  Marie-Aline Martin-Drumel for a machine-readable copy of her \CTHREE\ infrared spectrum covering the
frequency range 1900 - 2100~\cmi. This work used the DiRAC Data Intensive service (CSD3) at the University of Cambridge, managed by the University of Cambridge University Information Services  and the DiRAC Data Intensive service (DIaL2) at the University of Leicester, managed by the University of Leicester Research Computing Service on behalf of the STFC DiRAC HPC Facility (\url{www.dirac.ac.uk}). The DiRAC services at Cambridge and  Leicester were funded by BEIS, UKRI and STFC capital funding and STFC operations grants.
Computing facilities at University College London and the University of Oxford were also used.  This work was supported by the STFC Project No. ST/Y001508/1 and  by the European Research Council (ERC) under the European Union’s Horizon 2020 research and innovation programme through Advance Grant number 883830.  NFZ also acknowledges support by State Project IAP RAS No. FFUF-2024-0016.

\section*{Data Availability}

As with other ExoMol line lists, our energy levels (states file),
line lists in the form of Einstein A-Coefficients (trans file), 
calculated partition functions and opacities are available in the
ExoMol Database (\url{www.exomol.com}).
Our refined version of the \citet{2016JChPh.144d4307S} PES for
\CCC\ has been included in tabular form, along with a
{\scriptsize FORTRAN90} subroutine for reading it,
as supplementary material. 

\section*{Supplementary Material}

\cite{2016JChPh.144d4307S}'s PES has been
refined using experimental energy levels through the MARVEL procedure.
\citet[their table 3]{2016JChPh.144d4307S} provide non-redundant
parameters which, when used with their equation 3.1, defines the PES
they publish. Lines $4-61$ of the file

\begin{verbatim}
  C3_pes_refined.inp
\end{verbatim}

list our revisions to their non-redundant parameters; in the preceding lines,
the number of data records, equilibrium bond-lengths in \AA\ and the
equilibrium bond-angle supplement in degrees, have been provided.
With the additional file

\begin{verbatim}
  C3_pes.f90
\end{verbatim}

we have presented a {\scriptsize FORTRAN90} program which may be compiled
and used following

\begin{verbatim}
gfortran -o C3_pes.x C3_pes.f90
./C3_pes.x < C3_pes_refined.input
\end{verbatim}

to compute potential energies in \cmi\ for the bond-angles and
bond-lengths listed in Lines $62-72$ of the input file;
\XIICTHREE\ potential energies listed in the same lines should
agree with those computed to the digit.

\bibliographystyle{mnras}




\bibliography{}






\bsp	
\label{lastpage}
\end{document}